\documentclass[prl,showpacs,preprint,superscriptaddress]{revtex4}
\usepackage{amsfonts,amssymb,amsmath}
\usepackage{graphicx}
\usepackage{pstricks}

\newcommand{\be}{\begin{equation}}
\newcommand{\bea}{\begin{eqnarray}}
\newcommand{\ee}{\end{equation}}
\newcommand{\eea}{\end{eqnarray}}
\newcommand{\bpi}{\begin{picture}}
\newcommand{\bce}{\begin{center}}
\newcommand{\epi}{\end{picture}}
\newcommand{\ece}{\end{center}}

\begin{document}

\title{Gauge-invariant truncation scheme\\  
for the Schwinger-Dyson equations of QCD}
\date{December 17, 2007}

\author{D.~Binosi}
\email{binosi@ect.it}
\affiliation{ECT* - European Centre for Theoretical Studies in Nuclear
  Physics and Related Areas, Villa Tambosi, Strada delle
  Tabarelle 286, I-38050 Villazzano (TN), Italy.}
  
\author{J. Papavassiliou}
\email{joannis.papavassiliou@uv.es}
\affiliation{Departamento de F\'\i sica Te\'orica and IFIC, Centro Mixto, 
Universidad de Valencia-CSIC,
E-46100, Burjassot, Valencia, Spain,}

\begin{abstract}

We present  a new truncation scheme for  the Schwinger-Dyson equations
of QCD that respects gauge invariance at any level of the dressed loop
expansion. When applied to the gluon self-energy, it allows 
for its non-perturbative treatment without
compromising the transversality of the solution, even when entire sets of diagrams (most notably the ghost loops)
are omitted, or treated perturbatively.

\end{abstract}


\maketitle

{\it Introduction} -- The quantitative  understanding of the  
non-perturbative properties of
Quantum  Chromodynamics  (QCD) \cite{Marciano:1977su} constitutes   still  one  of  the  most
challenging problems  in particle physics.  The  basic building blocks
of  this  theory  are  the  Green's  (correlation)  functions  of  the
fundamental  degrees of  freedom,  gluons, quarks,  and ghosts.  
Their non-perturbative 
structure is at the center stage of extensive research that 
could furnish invaluable  clues for deciphering  the infrared
dynamics of QCD.

Lattice  simulations are  indispensable  in this  quest, since they 
capture all the non-perturbative  information of the theory. 
It  has become clear  by now that  the lattice
simulations yield  an infrared finite  gluon propagator in  the Landau
gauge.  This rather characteristic  behavior has recently  been firmly  established using
lattices  with  large  volumes; 
in addition,  the non-perturbative ghost propagator in  the same gauge
diverges,  at a  rate that  deviates only  mildly from  the tree-level
expectation~\cite{Bogolubsky:2007ud}.   These  clean  lattice  results  constitute  a  serious
challenge for the  QCD theorists: obtaining the same  results from the
theory formulated  in the continuum  is bound to expose  a fundamental
dynamical mechanism at work.

In the continuous formulation the  dynamics of all Green's functions 
are determined by 
 an infinite system  of coupled non-linear
integral equations 
known      as      Schwinger-Dyson      equations
(SDE)~\cite{Dyson:1949ha}.     These
equations  are inherently non-perturbative and
can be used  to address problems related to {\it e.g.},  chiral symmetry breaking,
dynamical  mass  generation,  and formation  of bound  states.   
Since this  system  involves an  infinite
hierarchy of equations,  in practice one is severely  limited in their
use, and the need for  a self-consistent truncation scheme is evident \cite{Curtis:1990zs}.
Devising such  a scheme, however,  is very challenging,  especially in
the  context of  non-abelian gauge  theories, like  QCD \cite{Mandelstam:1979xd}.   The central
problem stems from the fact that the SDEs are built out of unphysical
off-shell Green's functions; thus, the extraction of reliable physical
information  depends crucially  on  delicate all-order  cancellations,
which may be inadvertently distorted in the process of the truncation.

\begin{figure*}[!t]
\includegraphics[width=16.5cm]{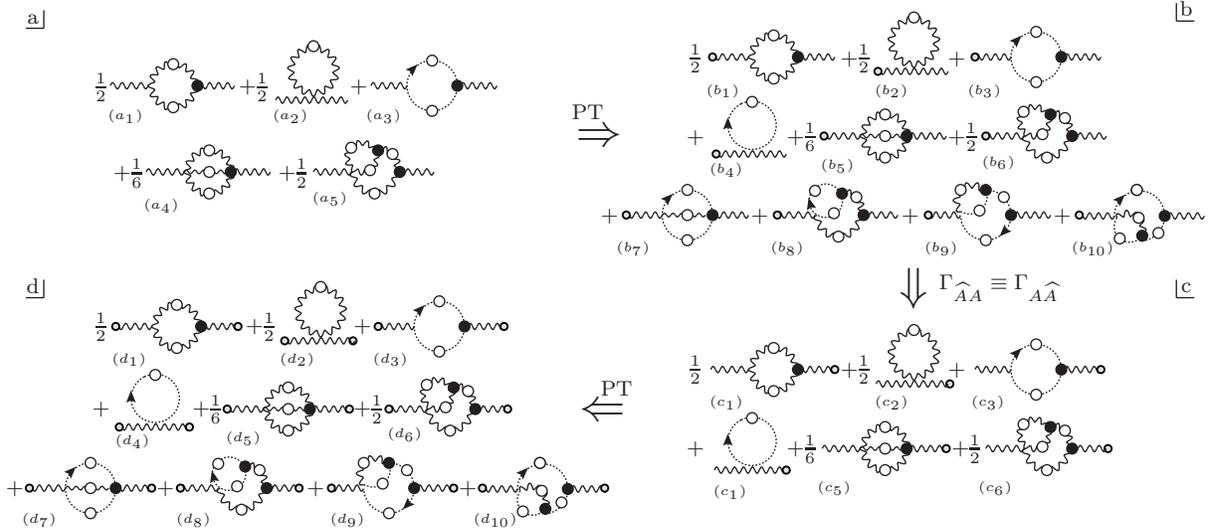}
\caption{The PT procedure to construct the new SDE of the gluon propagator. External legs ending in a gray circle represents background gluons.The corresponding Feynman rules can be found in \cite{Abbott:1980hw}.}
\label{fig:4steps}
\end{figure*}

The situation may best exemplified with the SDE of the 
gluon propagator $\Delta_{\alpha\beta}(q)$. 
In the Feynman gauge,  
\be
\Delta_{\alpha\beta}(q)=  -i\left[\left(g_{\alpha\beta} - \frac{q_\alpha q_\beta}{q^2}\right)\Delta(q^2) +\frac{q_\alpha
q_\beta}{q^4}\right],
\label{prop_cov}
\ee
where $\Pi_{\alpha\beta}(q)=(g_{\alpha\beta} - q_\alpha q_\beta/q^2) \Pi(q^2)$
is the gluon self-energy and
$\Delta^{-1}(q^2) = q^2 + i \Pi(q^2)$.
The conventional SDE for  $\Pi_{\mu\nu}$ reads
\be
 \Pi_{\alpha\beta}(q)=\sum_{i=1}^5(a_i)_{\alpha\beta},
 \label{SDE}
\ee
where the diagrams $(a_i)$ are shown in Fig.\ref{fig:4steps}a. 
Since the self-energy enters in the latter diagrams (white blobs in the same figure), Eq.(\ref{SDE})
constitutes a dynamical equation that can in principle determine $\Pi_{\alpha\beta}$.
Due to  general  arguments  based  on the  Becchi-Rouet-Stora-Tyutin (BRST)
symmetry~\cite{Becchi:1976nq},           $\Pi_{\alpha\beta}(q)$          is          transverse,
{\it i.e.}   $q^{\alpha}\Pi_{\alpha\beta}(q)  =0$.    Notice,   however,  that   
enforcing this fundamental property on the rhs of  Eq.(\ref{SDE}), {\it i.e.}, through
the contraction of  individual graphs by $q^{\alpha}$, is
far from  trivial, essentially  due to the  complicated Slavnov-Taylor
identities (STI) satisfied by  the fully-dressed vertices.  
As a result,  the
SDE of  Fig.\ref{fig:4steps}a cannot  be truncated without  compromising the
transversality of $\Pi_{\alpha\beta}(q)$.  For example, keeping only graphs
$(a_1)$ and $(a_2)$  is not correct even at  one loop.  Adding $(a_3)$
is still not sufficient for a SDE analysis, because (beyond one-loop)
$q^{\alpha}[(a_1)+(a_2) + (a_3)]_{\alpha\beta} \neq 0$.

In this letter we present a  new truncation scheme for the SDE of (quarkless) QCD
that  respects gauge  invariance  at  any level  of  the dressed  loop
expansion.  This becomes  possible  due to  the drastic  modifications
implemented  to  the building  blocks  of  the  SD series,  {\it i.e.}   the
off-shell  Green's  functions,  following the  field-theoretic  method
known as  pinch technique (PT)~\cite{Cornwall:1982zr}.
The PT  is a well-defined  algorithm that exploits  systematically the
BRST symmetry in order to construct new Green's functions endowed with
very  special  properties.  Most  importantly,  they satisfy  abelian,
Ward identities (WI) instead  of the usual STIs, have correct
analytic properties and displays only physical thresholds~\cite{Papavassiliou:1995fq}.

The PT rearrangement
gives rise {\it dynamically} to a new SD series analogous to the one in Eq.(\ref{SDE}), 
with the following characteristics: the graphs appearing on the rhs  
are made out of new vertices (Fig.\ref{fig:4steps}d), 
but contain the conventional self-energy $\Pi_{\alpha\beta}$ as 
before. These new vertices correspond 
precisely to the Feynman rules of the Background Field Method (BFM) in the Feynman gauge, {\it i.e.},
it is as if the external gluon had been converted 
dynamically into a background gluon.
The lhs, in addition to the term $\Pi_{\alpha\beta}(q)$ already there, contains 
additional terms, also proportional to $\Pi_{\alpha\beta}(q)$, which are generated during the PT 
rearrangement of the original rhs of Eq.(\ref{SDE}).

\begin{figure*}
\includegraphics[width=14cm]{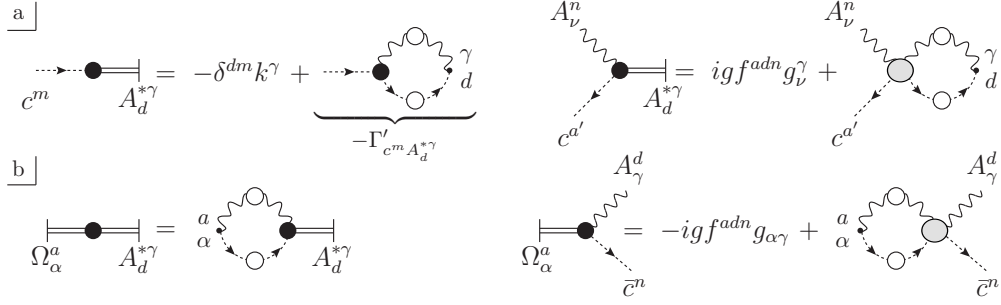}
\caption{The auxiliary functions $-\Gamma_{c^m A^{*\gamma}_d}$, $-\Gamma_{\Omega^a_\alpha A^{*\gamma}_d}$, $i\Gamma_{c^{a'} A^n_\nu A^{*\gamma}_d}$ and $i\Gamma_{\Omega^a_\alpha A^d_\gamma \bar c^n}$. Black and white blobs represent one-particle irreducible and connected Green's functions, respectively, while gray blobs are connected kernels.}
\label{fig:auxiliary}
\end{figure*}

{\it  A new  SD equation  for  the gluon  propagator} --  The
relevant  PT rearrangements take place when the longitudinal momenta 
of the three-gluon vertex trigger 
the  STIs satisfied by specific subsets  of fully dressed vertices
appearing  in  the  ordinary  perturbative
expansion.  Unlike   QED,  due  to  the  non-linearity   of  the  BRST
transformations,  these  STIs  are  realized through  auxiliary  (ghost)
Green's  functions involving  composite operators  such as  $\langle 0
\vert T[s\Phi(x)\cdots\vert0\rangle$,  where $s$ is  the BRST operator
and  $\Phi$ is  a  generic QCD  field.   It turns  out  that the  most
efficient  framework for  dealing with  these type  of objects  is the
so-called Batalin-Vilkovisky formalism~\cite{Batalin:1984jr}. In this framework, 
one adds to the original
gauge-invariant  Lagrangian  ${\cal  L}_\mathrm{I}$  the  term  ${\cal
L}_\mathrm{BRST}=\sum_\Phi\Phi^*s\Phi$,    coupling    the   composite
operators  $s\Phi$ to  the  BRST invariant  external sources (usually  called anti-fields)  $\Phi^*$, 
to  obtain the  new Lagrangian  ${\cal
L}_\mathrm{BV}={\cal   L}_\mathrm{I}+{\cal   L}_\mathrm{BRST}$.    One
advantage of  this formulation  is that it allows one to express the
STIs  of  the theory  in  terms of  auxiliary  functions  which can  be
constructed using  a well-defined  set of Feynman rules  (derived from
${\cal L}_\mathrm{BRST}$).   In particular, the usual  STI satisfied by
the  three-gluon  vertex,  an  essential  ingredient  in  the  ensuing
construction, assumes the form
\bea
q^\alpha\Gamma_{A^a_\alpha  A^m_\mu A^n_\nu}(k_1,k_2)&=&
q^2D^{aa'}(q)\left[\Gamma_{c^{a'} A^n_\nu A^{*\gamma}_d}(k_2,k_1)\Gamma_{A^d_\gamma A^m_\mu}(k_1)
\Gamma_{c^{a'} A^m_\mu A^{*\gamma}_d}(k_1,k_2)\Gamma_{A^d_\gamma A^n_\nu}(k_2)\right],\qquad
\label{STI:ggg}
\eea
where $\Gamma_{A^a_\alpha
A^b_\beta}(q)=(\Delta^{-1})^{ab}_{\alpha\beta}(q)-i\delta^{ab}q_\alpha    q_\beta$ [with $-\Gamma_{A_\alpha
A_\beta}(q)=\Pi_{\alpha\beta}(q)$],
and  the auxiliary  function  $\Gamma_{cAA^*}$, given  in Fig.\ref{fig:auxiliary}a,  is
nothing  but  the  standard  function appearing  in  the  conventional
derivation \cite{Ball:1980ax}  now written in the  anti-field language. An
important  property   of  auxiliary  functions   involving  the  gluon
anti-field, $A^*$, is encoded  into the so-called Faddeev-Popov equation:
\mbox{$\frac{\delta\Gamma}{\delta\bar
c^a}+iq^\mu\frac{\delta\Gamma}{\delta  A^{*a}_\mu}=0.$}  This equation
amounts to  the simple statement that contracting 
$A^*$ with its own momentum  $q$ converts it to an anti-ghost, $\bar
c$. This property will be used extensively in what follows.
In addition, one can obtain a set of useful identities relating Green's
functions  of background  fields  to those  of  quantum fields.  These
Background  Quantum Identities (BQIs)~\cite{Binosi:2002ez} are realized  through auxiliary
functions involving normal fields, anti-fields, and a background source
$\Omega$,  coupled  through  the  term  $-gf^{amn}\bar  c^a\Omega^m_\mu
A^\mu_n$, see Fig.\ref{fig:auxiliary}b.  The BQIs satisfied by the gluon propagator are
\bea                                      
i\Gamma_{\widehat{A}_\alpha^a
A_\beta^b}(q)&=&\left[ig_\alpha^\gamma                     
\delta^{ad}+
\Gamma_{\Omega_\alpha^a      A^{*\gamma}_d}(q)\right]\Gamma_{A^d_\gamma
A^b_\beta}(q),
\label{twoBQI1}\\
i\Gamma_{\widehat{A}_\alpha^a\widehat{A}_\beta^b}(q)&=&\left[i g_\alpha^\gamma
\delta^{ad}+ 
\Gamma_{\Omega_\alpha^a A^{*\gamma}_d}(q)\right]\Gamma_{A^d_\gamma
\widehat A^b_\beta}(q),
\label{twoBQI2}
\eea
which can be combined into the single identity
\bea
i\Gamma_{\widehat{A}_\alpha^a\widehat{A}_\beta^b}&=&i\Gamma_{A^a_\alpha A^b_\beta}+\Gamma_{\Omega_\alpha^a A^{*\gamma}_d}\Gamma_{A^d_\gamma A^b_\beta}+\Gamma_{\Omega_\beta^b A^{*\gamma}_d}\Gamma_{A^a_\alpha A^d_\gamma}\nonumber \\
&+&\Gamma_{\Omega_\alpha^a A^{*\gamma}_d}\Gamma_{A^d_\gamma A^{e}_{\epsilon}}\Gamma_{\Omega_\beta^b A^{*\epsilon}_{e}}.
\label{BQI:gg}
\eea
Other BQIs needed in our construction will be
\bea
i\Gamma_{\widehat{A}_\alpha^a\varphi  \phi}(k_1,k_2) &=& \left[
ig_\alpha^\gamma \delta^{ad} + \Gamma_{\Omega_\alpha^a A^{*\gamma}_d}(q)\right]
\Gamma_{\varphi A_\gamma^d \phi}(k_1,k_2) \nonumber \\
&+& R_{\Omega^a_\alpha\varphi  \phi}(k_1,k_2),
\label{genericBQI}
\eea
where $(\varphi,\phi)\in\{(A,A), (c, \bar{c}), (c, A^{*})\}$, and 
\bea
R_{\Omega^a_\alpha A^m_\mu A^n_\nu} &=& \Gamma_{\Omega^a_\alpha A^n_\nu A^{*\gamma}_d}
\Gamma_{A^d_\gamma A^m_\mu}+\Gamma_{\Omega^a_\alpha A^m_\mu A^{*\gamma}_d}
\Gamma_{A^d_\gamma A^n_\nu},\nonumber \\
R_{\Omega^a_\alpha c^m\bar{c}^n} &=& -\Gamma_{c^mA^{*\gamma}_{d}}\Gamma_{\Omega_\alpha^a A^d_\gamma \bar{c}^n} - 
\Gamma_{\Omega_\mu^a c^m c^{*d}}\Gamma_{c^d  \bar{c}^n},\\
R_{\Omega^a_\alpha c^m A^{*n}_\nu} &=&-\Gamma_{c^mA^{*\gamma}_{d}}\Gamma_{\Omega_\alpha^a A^d_\gamma A^{*n}_\nu}
-\Gamma_{\Omega_\alpha^a c^m c^{*d}}\Gamma_{c^dA^{*n}_\nu}.\nonumber 
\eea
Equipped with these relations we may now proceed to the derivation of our main result. 
The aim will be to start from the conventional SDE of Fig.\ref{fig:4steps}a and generate dynamically through the PT algorithm all the terms appearing in the BQI of Eq.(\ref{BQI:gg}), thus arriving at the SDE equation of Fig.\ref{fig:4steps}d. This will be accomplished by 
constructing the two BQIs of Eq.(\ref{twoBQI1}) and (\ref{twoBQI2}), one at a time.
The starting point is diagram $(a_1)$ of Fig.\ref{fig:4steps}a.
The tree-level three gluon vertex $\Gamma$ can be decomposed~\cite{Cornwall:1982zr}  into the sum $\Gamma^\mathrm{F}+\Gamma^\mathrm{P}$, where (factoring out the color structure)
\bea
i\Gamma^\mathrm{F}_{A_\alpha A_\mu A_\nu}(k_1,k_2)&=&g_{\mu\nu}(k_1-k_2)_\alpha-2q_\mu g_{\alpha\nu}+2q_\nu g_{\alpha\mu},
\nonumber\\
i\Gamma^\mathrm{P}_{A_\alpha A_\mu A_\nu}(k_1,k_2)&=&g_{\alpha\nu}k_{1\mu}-g_{\alpha\mu}k_{2\nu}.
\label{PT_splitting}
\eea
This splitting assigns a special role to the physical momentum $q$, making $\Gamma^\mathrm{F}$ 
Bose symmetric only with respect to the $A_\mu$ and $A_\nu$ legs inside the loop. In fact,
$\Gamma^\mathrm{F}$     coincides      with     the     BFM     vertex
$\Gamma^{(0)}_{\widehat{A}_\alpha  A_\mu A_\nu}$. 
$\Gamma^\mathrm{P}$ contains  the  longitudinal  momenta that
will get  contracted with the full three-gluon  vertex, triggering the
STI     of     Eq.(\ref{STI:ggg}).      The     result     will     be
$(a_1)=(a_1)^\mathrm{F}+(a_1)^\mathrm{P}$,   with   $(a_1)^\mathrm{F}$
coinciding with diagram $(b_1)$, and
\bea
(a_1^\mathrm{P})^{ab}_{\alpha\beta}&=&-i\Gamma_{\Omega_\alpha^a A^{*\gamma}_d}(q)\Gamma_{A^d_\gamma A^b_\beta}(q)-igf^{amd}\left\{\int_{k_1}k_{2\alpha}D(k_1)D(k_2)\Gamma_{c^m A^b_\beta \bar c^d}(-q,k_2)\right.\nonumber \\
&+&\left.\int_{k_1}D(k_1)\Gamma'_{c^e A^{*d}_\alpha}(k_2)D(k_2)\Gamma_{c^m A^b_\beta\bar c^e}(-q,k_2)+i\int_{k_1}D(k_1)\Gamma_{c^mA^b_\beta A^{*d}_\alpha}(-q,k_2)\right\}.\qquad
\label{first_step}
\eea
In the equation above we have used the ghost SDE \mbox{$k^2D(k)=1-i\Gamma_{c \bar
c}(k)D(k)$} to  transform a  tree-level ghost propagator  appearing in
the  second term of the rhs  into a  full one.  The first
integral   on  the  rhs   of  Eq.(\ref{first_step})   symmetrizes  the
ghost-gluon vertex  of $(a_3)$, giving  rise to the  characteristic BFM
vertex $\propto  (k_1-k_2)_\alpha$, and thus to  diagram $(b_3)$.  The
second term coincides precisely with the diagram $(b_{10})$; the third
term  (see  Fig.\ref{fig:auxiliary})  gives  rise  to  diagram  $(b_4)$  [through  the
tree-level part  of $\Gamma_{cAA^*}$], as well as  $(b_6)$, $(b_7)$ and
$(b_8)$. Finally,  due to the fact  that the four gluon
vertices  $\Gamma_{\widehat AAAA}$  and  $\Gamma_{AAAA}$ coincide at tree-level,  we
will have $(a_2)=(b_2)$, $(a_4)=(b_5)$, and $(a_5)=(b_6)$.  Thus taking
into  account   the  first  term  in   Eq.(\ref{first_step})  we  have
dynamically reproduced the propagator BQI of Eq.(\ref{twoBQI1}).

At  this  point   we  have  constructed  $\Gamma_{\widehat  A^a_\alpha
 A^b_\beta}(q)$; the next step will be to exploit the obvious equality
 $\Gamma_{\widehat     A^a_\alpha     A^b_\beta}(q)=\Gamma_{A^a_\alpha
 \widehat  A^b_\beta}(q)$ to  interchange the  background  and quantum
 legs (see Fig.\ref{fig:4steps}c). This introduces a considerable simplification: on
 the  one hand  we  keep  identifying the  pinching  momenta from  the
 the PT decomposition of the (tree-level) $\Gamma$,  while on the  other hand the  equality between
 diagrams $(c_5)$, $(c_6)$ and $(d_5)$, $(d_6)$ is immediate.

Let us now carry out the PT splitting of Eq.(\ref{PT_splitting}) to diagram $(c_1)$. The $\Gamma^\mathrm{F}$ part of the vertex generates directly diagram $(d_1)$; the longitudinal momenta contained in $\Gamma^\mathrm{P}$ get contracted as before with the full three-gluon vertex, which, however, has now an external background leg.
Using Eq.(\ref{genericBQI}) with $(\varphi,\phi)=(A,A)$, we get
\bea
(c_1^\mathrm{P})^{ab}_{\alpha\beta}&=&-i\left[
ig_\beta^\gamma \delta^{bd} + \Gamma_{\Omega_\beta^b A^{*\gamma}_d}(q)\right](a_1^\mathrm{P})^{ad}_{\alpha\gamma}+gf^{amn}\int_{k_1}\!\!\Delta^\nu_\alpha(k_2)\frac{k_1^\mu}{k_1^2} R_{\Omega^b_\beta A^m_\mu A^n_\nu}(k_1,k_2).\quad
\label{second_step}
\eea The presence  of the prefactor \mbox{$ig_\beta^\gamma \delta^{bd}
+ \Gamma_{\Omega_\beta^b  A^{*\gamma}_d}$} allows one to  use the BQIs
of Eq.(\ref{genericBQI}) to  convert the full vertices $\Gamma_{cA\bar
c}$  and  $\Gamma_{cAA^* }$,  appearing  in  the  last three  terms  of
$(a_1^\mathrm{P})$,    into     $\Gamma_{c\widehat{A}\bar    c}$    and
$\Gamma_{c\widehat{A}A^*  }$, respectively.   This  operation has  two
effects:  ({\it  i})  it  generates  $(d_7)$,  $(d_8)$,  $(d_9)$,  and
$(d_{10})$,  plus  the  contribution  needed to  convert  $(c_3)$  and
$(c_4)$ into  $(d_3)$ and ($d_4$),  respectively; ({\it ii})  it gives
rise to leftover contributions  given by the three integrals appearing
in Eq.(\ref{first_step}) where the corresponding vertex is replaced by
either  $-R_{\Omega c\bar  c}$  or $-R_{\Omega  cA^*}$.  These  latter
terms cancel exactly against the second term in Eq.(\ref{second_step}),
after  its tree-level  contribution  has been  extracted  and used  to
convert  ($c_2$) into~($d_2$).  At this  point we  have  generated all
diagrams of Fig.\ref{fig:4steps}d~\cite{Sohn:1985em}. 
 In  addition, using the BQI of Eq.(\ref{twoBQI1})
(already proven  in the previous step), the  term in $(a_1^\mathrm{P})$
proportional   to   $\Gamma_{\Omega    A^*}$   will   give   precisely
$-i\Gamma_{\Omega_\alpha^a A^{*\gamma}_d}(q)\Gamma_{A^d_\gamma\widehat
A^b_\beta}(q)$.  Thus, we  have constructed  the full BQI of
Eq.(\ref{twoBQI2}).    Having  dynamically   realized   the  BQIs   of
Eqs.(\ref{twoBQI1})  and  (\ref{twoBQI2}), we  can  combine them  into
Eq.(\ref{BQI:gg}), which constitutes the announced result.
We emphasize that ({\it i}) all rearrangements have been induced by the PT
manipulation of only one diagram [$(a_1)$ and $(c_1)$ of Fig.\ref{fig:4steps}]
and ({\it ii}) all quantities encountered  exist in the conventional formulation. 
In that sense, the Batalin-Vilkovisky formalism serves simply as an efficient 
way of keeping track of them.

{\it Discussion} -- The new SD series just constructed reads
\be
 [1+G(q^2)]^2\Pi_{\alpha\beta}(q)= \sum_{i=1}^{10}(d_i)_{\alpha\beta},
\label{SDE1}
\ee
where $G$ is defined as the part of $\Gamma_{\Omega_\alpha A^*_\beta}$ proportional to  
$g_{\alpha\beta}$: $\Gamma_{\Omega_\alpha A^*_\beta}(q)=iG(q^2)g_{\alpha\beta}+\sim q_\alpha    q_\beta$. The rhs of Eq.(\ref{SDE1}) has a very special structure. 
The diagrams of Fig.\ref{fig:4steps}d can be separated into four subgroups
[$(d_1)$, $(d_2)$], [$(d_3)$, $(d_4)$], [$(d_5)$, $(d_6)$], and [$(d_7)$, $(d_8)$, $(d_9)$, $(d_{10})$],
corresponding to one- or two-loop dressed gluonic or ghost contributions.
Due to  the abelian WIs satisfied by these new vertices,
the contribution of each of the four subgroups  
is {\it individually} transverse \cite{Aguilar:2006gr}.

The practical implications of this property for the 
treatment of the SD series are far-reaching, 
since it furnishes 
a systematic, manifestly gauge-invariant truncation scheme.
In the case of the gluon self-energy, for instance, 
the transversality of the answer is guaranteed at every step.
Specifically, keeping only the 
diagrams in the first group, we obtain the truncated SDE
\be
\Pi_{\alpha\beta}(q) = [1+G(q^2)]^{-2}
[(d_1)+(d_2)]_{\alpha\beta},
\label{trua}
\ee
and we have that 
$q^{\alpha}[(d_1)+(d_2)]_{\alpha\beta}=0$
by virtue of $q^{\alpha}{\Gamma}_{\widehat{A}^a_\alpha A^m_\mu A^n_\nu}(k_1,k_2)=
gf^{amn}[\Delta^{-1}_{\mu\nu}(k_1)-\Delta^{-1}_{\mu\nu}(k_2)]$. Therefore, 
 $\Pi_{\alpha\beta}(q)$ is transverse, as it should, despite the omission of 
the remaining graphs (most notably the ghost loops). In fact, one can envisage the possibility of employing  
completely different treatments for each subgroup: for example, one may treat the graphs $(d_1)$ and $(d_2)$ 
non-perturbatively, while opting for  a perturbative treatment of the ghost diagrams $(d_3)$ and $(d_4)$, without compromising the  
transversality of the self-energy. The price one has to pay is the need to consider the additional SDE governing $G$ (see Fig.\ref{fig:auxiliary}b). Notice, however, that the approximations employed for the treatment of this latter SDE will not interfere with the transversality of $\Pi_{\alpha\beta}$.
The abelian WIs furnish an additional technical advantage:
one may use gauge-technique inspired Ans\"atze, a common practice when
dealing  with the  SDE of  QED~\cite{Curtis:1990zs}, to  express the  vertices in  terms of
propagators,  in  such  a   way  as  to  automatically  enforce  gauge
invariance.
Finally, notice that ({\it i}) the SDEs for the QCD vertices can be constructed in a very similar way~\cite{quesera}, and ({\it ii}) 
the analysis presented here can be generalized to other gauges ({\it e.g.}, the Landau gauge) using the 
methodology developed in~\cite{Pilaftsis:1996fh}.

In conclusion, the new SD series constructed in this letter provides a powerful tool for the 
systematic exploration of the non-perturbative sector of QCD, allowing 
the study of the fundamental Green's functions in a manifestly gauge-invariant way.

{\it Acknowledgments:} DB thanks the Physics Department of the University of Valencia, 
where part of this work has been carried out.
JP is supported by the MEC grant FPA 2005-01678 
and the Fundaci\'on General of the UV. Diagrams drawn using JaxoDraw~\cite{Binosi:2003yf}.

\end{document}